\documentclass[runningheads,a4paper]{llncs}

\usepackage[english]{babel}
\usepackage{lipsum}
\usepackage{graphicx}
\usepackage{caption}
\usepackage{subcaption}
\usepackage{floatrow}
\usepackage[hidelinks]{hyperref}
\usepackage[space]{cite}
\usepackage[usenames,dvipsnames]{color}
\usepackage{listings}
\usepackage{booktabs}
\usepackage{empheq}
\usepackage{xspace}
\usepackage[ruled, vlined, linesnumbered,algosection,algo2e]{algorithm2e}
\SetAlFnt{\ssmall} %
\usepackage{moresize}
\usepackage{paralist}
\usepackage{pifont}
\usepackage{footmisc}%

\usepackage{float}
\floatstyle{plaintop}
\restylefloat{table}

\begin{document}

\author{Patrick Wollgast, Robert Gawlik, Behrad Garmany, \\ Benjamin Kollenda, Thorsten Holz}
\authorrunning{Wollgast et al.}
\authorrunning{Wollgast, Gawlik, Garmany, Kollenda and Holz}
\title{Automated Multi-Architectural Discovery of CFI-Resistant Code Gadgets}
\institute{Horst G\"ortz Institute for IT-Security (HGI)\\ Ruhr-Universit\"at Bochum, Germany} %

\maketitle

\begin{abstract}

Memory corruption vulnerabilities are still a severe threat for software systems. To thwart
the exploitation of such vulnerabilities, many different kinds of defenses have been proposed in the past. Most prominently, \emph{Control-Flow Integrity} (CFI) has received a lot of attention recently. Several proposals were published that apply coarse-grained policies with a low performance overhead. However, their security remains questionable as recent attacks have shown. 

To ease the assessment of a given CFI implementation, we introduce a framework to discover code gadgets for code-reuse attacks that conform to coarse-grained CFI policies. 
For this purpose, binary code is extracted and transformed to a symbolic representation in an
architecture-independent manner. Additionally, code gadgets are verified to
provide the needed functionality for a security researcher. We show that our
framework finds more CFI-compatible gadgets compared to other code gadget discovery
tools. Furthermore, we demonstrate that code gadgets needed to bypass CFI solutions on the
ARM architecture can be discovered by our framework as well.

\end{abstract}

\section{Introduction}

Memory corruption vulnerabilities have threatened software systems for decades.
The deployment of various defense mechanisms, such as  \emph{data execution prevention} (DEP)~\cite{dep}, \emph{stack smashing protection} 
(SSP)~\cite{Cowan:1998:SAA:1267549.1267554}, 
and \emph{address space layout randomization} (ASLR)~\cite{aslr} have raised the
bar for reliable memory corruption exploitation significantly.
Nevertheless, a dedicated attacker is still able to achieve code
execution~\cite{vupen1,vupen2}. \emph{Information leaks} are utilized to counter
ASLR and reveal the layout of the address space, or to harvest code to build
a payload just-in-time
~\cite{vupen1, Snow:2013:JCR:2497621.2498135}. To circumvent DEP, 
attackers have added code-reuse attacks to their repertoire, such as \emph{return-oriented
programming}
(ROP)~\cite{Shacham:2007:GIF:1315245.1315313,Buchanan:2008:GIG:1455770.1455776,kornau},
\emph{jump-oriented programming} (JOP)\
\cite{Davi10return-orientedprogramming,Checkoway:2010:RPW:1866307.1866370,Bletsch:2011:JPN:1966913.1966919},
and \emph{call-oriented programming} (COP)~\cite{184507}.
Code-reuse attacks do not inject new code but chain together small chunks of
existing code, called \emph{gadgets}, to achieve arbitrary code execution.

In response to this success, the defensive research was driven to find
protection methods against code-reuse attacks. Some results of this research are
\emph{kBouncer}~\cite{182945},
\emph{ROPecker}~\cite{DBLP:conf/ndss/ChengZYDD14},
\emph{EMET}~\cite{EMET} including \emph{ROPGuard}~\cite{EMETROPGuard},
\emph{BinCFI}~\cite{180373}, and
\emph{CCFIR}~\cite{Zhang:2013:PCF:2497621.2498134}. These defenses incorporate
two main ideas. The first is to enforce \emph{control-flow integrity} (CFI)
~\cite{Abadi:2005:CI:1102120.1102165,Abadi:2009:CIP:1609956.1609960}. With
perfect CFI, the control-flow can neither be hijacked by code-injection nor by
code-reuse~\cite{Goktas:2014:OCO:2650286.2650770}. However, the overhead of
perfect CFI is too high to be practical. Therefore, the proposed defense methods
try to strike a balance between security and tolerable overhead. The second
idea is to detect code-reuse attacks by known characteristics of an attack like
a certain amount of gadgets chained together. All of those schemes defend
attacks on the x86/x86-64 architecture. For other architectures the research is
lacking behind~\cite{DBLP:conf/ndss/DaviDEFHHNS12,pewny2013control}.
Several generic attack vectors have been published by the offensive side to
highlight the limitations of the proposed defense methods. Although single
implementations can be bypassed with common code-reuse attacks by exploiting a
vulnerability in the implementation~\cite{DaBEMET,BMEMET}, generic
circumventions rely
on longer and more complex
gadgets~\cite{Goktas:2014:OCO:2650286.2650770,184481,DBLP:conf/raid/SchusterTPMSCH14,184507,184515}
or complete functions~\cite{TUD-CS-2015-0041}.
Since the gadgets loose their simplicity by becoming longer, it also becomes
harder to find specific gadgets and chain them together. 
To the best of our knowledge there is no gadget discovery framework available
to search for CFI resistant gadgets.
To be able to assess a CFI solution, it is necessary to discover code gadgets
which could execute within the boundaries of the solution's CFI policies or
detection heuristics. We
provide a framework which is able to discover CFI resistant code gadgets or
complete functions across different architectures, an increasingly important property as CFI starts to evolve
on non-x86 systems as well. Notably, no search for
CFI resistant code gadgets has been performed for ARM, while defenses for this architecture have already been
developed~\cite{DBLP:conf/ndss/DaviDEFHHNS12, pewny2013control}.
The information provided by our framework helps security researchers to
quickly prototype exploit examples to test a given CFI solution.

We opted to use an \emph{intermediate language} (IL) for the analysis of
extracted code to support different architectures without the
effort to adjust the algorithms to new architectures. 
Because of the high architecture coverage, \emph{VEX} is our
choice for the IL. VEX is part of \emph{Valgrind}, an instrumentation framework
intended for dynamic use~\cite{val}. 
We harness VEX in static analysis manner~\cite{pyvex,
shoshitaishvili2015firmalice} and utilize the SMT solver
\emph{Z3}~\cite{Z3} to translate
code gadgets into a symbolic representation to enable symbolic execution and
path constraint analysis.
Our evaluations shows that our framework discovers 1.2 to 154.3 times more CFI-resistant gadgets across different architectures and operating systems than
other gadget discovery tools. Additionally, we show that CFI-resistant gadgets
are available in binary code for the ARM architecture as well, which should be
taken into account by future CFI solutions.

In summary, we make the following contributions:
\begin{itemize}
    \item We develop a framework to discover CFI and heuristic-check resistant gadgets in an architecture-independent, offline search.
    \item Our framework delivers semantic definitions of extracted code gadgets and
        classifies them based on these definitions for convenient search and
        utilization by a security researcher.
    \item To the best of our knowledge, we are the first to provide a code gadget
        discovery framework which reveals CFI resistant gadgets across different processor
        architectures, and show that CFI-compatible gadgets are also prevalent on the ARM architecture.
\end{itemize}

\section{Technical Background}

\label{background}

We begin by briefly describing code-reuse attacks, CFI approaches, and heuristic techniques proposed by recent research to defend against runtime attacks. It is important to understand the concept of CFI and the heuristic checks, as we focus on gadgets that are resistant against these approaches. Architecture independence is another issue that is tackled by our framework.

\subsection{Code-Reuse Attacks}
The introduction of \emph{data execution prevention} (DEP)~\cite{dep} on modern operating systems provided a useful protection against the injection of new code. 
To bypass DEP, attackers often resort to reusing 
code already provided by the vulnerable executable itself (or one of its 
libraries). Vulnerabilities suitable for code-reuse attacks are memory corruptions
such as stack, heap or integer overflows, or a dangling pointer. The technique 
most commonly applied to reuse existing code is \emph{return-oriented programming}
 (ROP)~\cite{Shacham:2007:GIF:1315245.1315313,Buchanan:2008:GIG:1455770.1455776}. 
 The concept behind ROP is to combine small sequences of code, called 
\emph{gadgets}, that end with a return instruction. 
All combined gadgets of an exploit are often referred to as a \emph{gadget chain}. 
To be able to combine these gadgets, either a sequence of return addresses has to be placed on the stack where each address points to the next gadget,
or the stack pointer has to be redirected to a buffer containing these addresses. The process of redirecting the stack 
pointer is called \emph{stack pivoting}. 

For architectures with variable opcode length like x86/x86-64, the instructions used for the gadgets do not have to be
aligned as intended by the compiler. Previous work has shown that enough gadgets for arbitrary computations can be 
located~\cite{kornau,Davi10return-orientedprogramming,Buchanan:2008:GIG:1455770.1455776} even without those unintended instructions. This is an interesting observation that especially concerns architectures with fixed opcode length. Automated tools that search for gadgets and chain them together have also been developed by past research~\cite{Hund:2009:RRB:1855768.1855792,ropc}.

Over the years, research on code-reuse attacks has proposed different variations of ROP such as \emph{jump-oriented programming} (JOP)
~\cite{Davi10return-orientedprogramming,Checkoway:2010:RPW:1866307.1866370,Bletsch:2011:JPN:1966913.1966919} and
\emph{call-oriented programming} (COP)~\cite{184507}. JOP uses jumps instead of returns to 
direct the control-flow to the next gadget, and COP uses calls. 
Due to their complexity, code-reuse attacks are typically used to make 
injected code executable thus defeating protections like DEP and redirect the
control-flow to the injected code~\cite{vupen1,vupen2}.

\subsection{Control-Flow Integrity (CFI)}
The concept of CFI was first introduced by Abadi et al.~\cite{Abadi:2005:CI:1102120.1102165,Abadi:2009:CIP:1609956.1609960}.
A program maintains the CFI property, if the control flow remains in a predefined \emph{control-flow graph} (CFG). This predefined CFG contains all intended execution paths of the program. 
If an attacker redirects the control flow via code injection 
or code-reuse attacks to an unintended execution path, the CFI property is violated and the attack is
detected. In an ideal CFG, every indirect transfer has a list of valid unique identifiers (IDs)
and every transfer target has an ID assigned to it~\cite{Goktas:2014:OCO:2650286.2650770}. These IDs are checked before indirect transfers occur to ensure that the target is valid.

If CFI is applied to proprietary software, it becomes problematic to generate such
a detailed CFG. To construct the CFG, the program has to be disassembled and a pointer 
analysis performed. Every error made during this process may lead to false 
positives during runtime of the protected program. Another issue with the classical CFI approach as proposed by Abadi et al. is performance. 
Therefore, implemented CFI solutions---also called coarse-grained approaches---typically reduce the number of IDs by assigning the same ID to the same
category of targets.
Examples of coarse-grained approaches are BinCFI~\cite{180373} and CCFIR~\cite{Zhang:2013:PCF:2497621.2498134}.
BinCFI uses two IDs to ensure the integrity of the CFG. The first ID defines rules for targets of \emph{return} (RET) instructions and \emph{indirect jumps} (IJ). 
The second ID combines rules for indirect control-transfers from the \emph{procedure linkage table} (PLT) and indirect calls (ICs). 
Each ID has its own routine which resides inside the protected binary. Every indirect transfer is instrumented to jump to one of the two verification routines. 
Similar to BinCFI, CCFIR is also a coarse-grained CFI approach applied to binaries without source code. 
Each indirect transfer is redirected through a \emph{Springboard}. The Springboard contains all valid control-flow targets and thereby prevents that the flow is redirected to invalid targets. An initial permutation of the Springboard at program startup additionally raises the bar for attackers. 

\subsection{Heuristic Approaches}
\label{heuristic_antirop}
In 2013 Pappas et al.~\cite{182945} introduced kBouncer, an heuristic-aided approach that leverages modern hardware features to prevent code-reuse attacks.
To perform CFI checks, kBouncer utilizes the \emph{Last Branch Record} (LBR). LBR is a feature of contemporary Intel and AMD processors which can only be enabled and disabled in kernel mode. Therefore, kBouncer consists of a user and kernel mode component. Like the name suggests, LBR records the last taken branches or a subset of the last branches. 
Each entry in the LBR contains the source and destination address of the taken branch. By fetching some bytes just before the destination address, kBouncer can examine and enforce that every return address is preceded by a call instruction. Otherwise kBouncer reports a CFI violation. 
Besides the CFI enforcement, a heuristic check is performed by inspecting the last 8 indirect branches. 
If all entries match kBouncers gadget definition, an attack is reported. A gadget is considered as an entry if it contains up to 20 instructions and
ends in an indirect control flow. 
The checks are invoked whenever one out of 52 critical WinAPI functions such as \emph{VirtualProtect} or \emph{WinExec} is called. The user-mode component hooks these critical functions and
triggers the checks in the kernel mode component. 

Another heuristic-aided approach is ROPecker by Cheng et al.~\cite{DBLP:conf/ndss/ChengZYDD14}, which also utilizes the LBR stack to look for gadgets in the past control flow. Additionally, the future control flow is also examined. To check for gadgets in the future control flow, ROPecker combines online emulation of the
flow, stack inspection, and an offline gadget search. Since gadgets are already searched
offline and stored to a database, ROPecker has also the possibility to detect unaligned
gadgets. 
To detect gadgets, ROPecker does not apply CFI enforcements, but merely relies on heuristics.
A gadget in the context of ROPecker is a sequence of up to 6 instructions ending with an indirect control-flow transfer. Sequences containing direct branch instructions are excluded from the definition. ROPecker inspects the past control flow first by utilizing the LBR to record indirect branch instructions. 
The first non-gadget encountered while walking the LBR backwards terminates the search for gadgets in the past
control flow. Afterwards, the future control flow is 
inspected for gadgets. If the combined number of encountered gadgets from the past and future control flow is above a predefined threshold, an attack is reported. The research of Cheng et al. suggests that a threshold 
between 11 and 16 gadgets is a suitable number.

\subsection{Defeating the Countermeasures}
All presented defenses against code-reuse attacks have been bypassed in recent years. While some attacks exploit vulnerabilities in a specific implementation to disable the checks~\cite{DaBEMET,BMEMET}, we focus on generic bypasses to defeat the protections. We divide the defense policies in two categories, CFI policies posing limitations on indirect branch instructions and heuristic policies looking for typical characteristics of code-reuse attack vectors.

\label{cfi_policies}
Attacks focusing on kBouncer, ROPecker, and EMET/ROPGuard~\cite{184515,184507,DBLP:conf/raid/SchusterTPMSCH14} just have to bypass the call site (CS) checks. However, attacks against BinCFI and CCFIR~\cite{Goktas:2014:OCO:2650286.2650770,184481} also have to take into account that ICs and IJs are limited to certain control-flow targets like function entry points (EPs). G\"{o}kta\c{s} et al.~\cite{Goktas:2014:OCO:2650286.2650770} categorize the gadgets by their prefix (CS or EP), their payload (IC, \emph{fixed function call} (F), other instructions), and their suffix (RET, IC, IJ). This categorization results in 18 (2 $\cdot$ 3 $\cdot$ 3) different gadget types.
They even use gadgets containing conditional jumps. With these gadget categories, they are able to bypass CCFIR, which they consider stricter than BinCFI. Another interesting gadget type is the \emph{i-loop-gadget}~\cite{DBLP:conf/raid/SchusterTPMSCH14}. In their work, Schuster et al. use a loop containing an IC to chain gadgets and invoke security sensitive functions.

The heuristic policies explained in \S~\ref{heuristic_antirop} check for chains of short instruction sequences. To evade these checks, long gadgets with minimal side effects were proposed~\cite{184515,184507}. If the heuristic check
encounters a long instruction sequence, the evaluation is terminated and the chain is classified as benign. Another elegant method is to invoke a function call to an unsuspicious function like \emph{lstrcmpiW}~\cite{DBLP:conf/raid/SchusterTPMSCH14}. If the unsuspicious function does not alter the global state of the program and takes enough indirect branches, the attack cannot be discovered by the heuristic checks.

\section{Design and Implementation}

The process of discovering suitable code gadgets which fulfill certain CFI
policies consist of broadly two phases: first, appropriate code has to be
discovered and extracted. Second, it is translated into the symbolic
representation and can then be classified according to semantic definitions.

\subsection{Gadget Discovery}
\label{gadget_disco}
Before we can describe the process of the gadget discovery, we have to define
the gadgets' properties first. The definition of the gadgets is important as they
define the bounds and specify the content of the gadgets. After the definition
of the gadgets is given, we introduce the algorithms to locate all points of
interest for the gadget discovery and the algorithm to discover the gadgets
themselves.
\paragraph{Gadget Categories.}
\label{gadget_categories}
Except minor modifications, our gadgets conform to the
specifications defined by by G\"{o}kta\c{s} et
al.~\cite{Goktas:2014:OCO:2650286.2650770} and Schuster et
al.~\cite{DBLP:conf/raid/SchusterTPMSCH14} as explained in \S~\ref{cfi_policies}.
Their definitions provide
sufficient properties to, for example, find complete functions for code-reuse and other
CFI resistant gadgets.
We used their definitions to restrict the gadget discovery, but definitions can
be extended and added in modular fashion to our framework to support additional gadget types.
The bounds of our gadgets have to conform to legitimate control-flow targets. Thus, they
have to start at an EP or at a CS and end with an IC, IJ, or RET.
The content of a gadget is defined as either an IC, a
\emph{fixed function call} (F), or other arbitrary instructions.
We opted to drop
IC as gadget content definition, because we can connect a gadget ending with an IC
with the gadget it follows starting at the CS. 
Fixed function calls are beneficial in two ways. Instead of
reading the address of the function from the \emph{import address table}
(IAT) and preparing the call, one can simply use the gadget with the fixed
function call. However, this just works if all parameters of the function can be
set to the desired values. Furthermore, defenses preventing calls to security
sensitive functions~\cite{Zhang:2013:PCF:2497621.2498134} can be circumvented by
using gadgets containing a legitimate call to the function.
As we show in ~\S~\ref{_sec:gadget_dist}, many hardcoded function calls inside of gadgets exist.

\begin{figure}[t]
\begin{floatrow}
\ffigbox{%
	\includegraphics[width=0.5\linewidth]{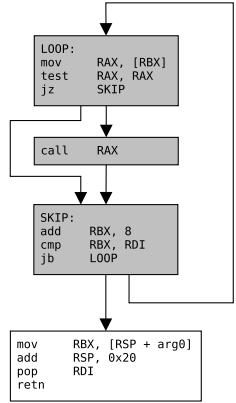}
}{%
 	\caption{Instructions of an example loop gadget.
    Just the gray basic blocks belong to a loop gadget by our definition.}	
	\label{fig:loop}
}
\capbtabbox{%
	\centering
    \ssmall
	\caption[All gadget definitions.]{Gadget types supported by our framework.}
	\label{tab_gadget_def}
	\begin{tabular}{ccc}
	 \\
	Prefix & Content & Suffix \\
	\midrule
	EP & Arbitrary Instructions & IC \\
	\midrule
	EP & Arbitrary Instructions & IJ \\
	\midrule
	EP & Arbitrary Instructions & RET \\
	\midrule
	EP & F & IC \\
	\midrule
	EP & F & IJ \\
	\midrule
	EP & F & RET \\
	\midrule
	CS & Arbitrary Instructions & IC \\
	\midrule
	CS & Arbitrary Instructions & IJ \\
	\midrule
	CS & Arbitrary Instructions & RET \\
	\midrule
	CS & F & IC \\
	\midrule
	CS & F & IJ \\
	\midrule
	CS & F & RET \\
	\midrule
	CS & Loop & IC \\
	\bottomrule
	\end{tabular}

}{%
}
\end{floatrow}
\end{figure}

Another useful gadget is the loop gadget. Loops can be used as a \emph{dispatch
gadget}~\cite{DBLP:conf/raid/SchusterTPMSCH14,Bletsch:2011:JPN:1966913.1966919}
to invoke other gadgets. Figure~\ref{fig:loop} shows a gadget proposed by
Schuster et al. During the first iteration of the loop, RBX points to the
beginning of a list with the addresses of the to-be dispatched gadgets. RDI
points to the end of this list during all iterations of the loop. 
If the end of the loop is reached
the gadget returns. The difference between the proposed gadget and
the gadget defined for our search is that just the gray basic blocks in Figure
\ref{fig:loop} belong to our loop gadget definition.
For simplicity, loop gadgets end with an IC and start either at the CS of its IC or at
an EP.
Hence, the basic block beginning with the
label \emph{@skip} and the last basic block comprise a separate, overlapping CS-RET gadget. This
has the advantage that also loop gadgets in big functions
without a tailing gadget (CS-RET) are found.
Additionally, one can query if another gadget starts at
the end of the loop gadget. This way, when searching for \emph{tailless}
loop gadgets, we can query if code which overlaps, comprises a gadget containing
another suffix than RET.
All supported gadget definitions are summarized in Table~\ref{tab_gadget_def}.
These definitions allow us to extract code with conditional jumps such that each single code path represents a
single gadget in a path-insensitive way. As each of them is verified with
symbolic execution later on, path-sensitive code gadgets arise and path-insensitive gadgets are dropped (see \S~\ref{gsma:ana}).

\newfloatcommand{capbtabbox}{table}[][\FBwidth]

\paragraph{Discovering Points of Interest.}
To locate gadgets, our search algorithm follows the paths of the CFG.
The starting points for the search algorithm are IC, IJ, and RET instructions.
The algorithm to locate these points of interest
works in two phases.
In the first phase, addresses of all calls to fixed functions in all modules
of a program of interest are extracted and kept. The set of fixed
functions comprises critical imported functions which handle memory management, process and
thread creation, and file I/O. These are typically very valuable for an
attacker.
During the second phase, the algorithm iterates over every instruction belonging
to a function.
If an instruction is a RET, IC, IJ, or a call, the address of the
instruction is added to the corresponding list of starting points.

\paragraph{Gadget Extraction with Depth-First Search.}
To retrieve the gadgets shown in Table~\ref{tab_gadget_def}, we have to traverse the
CFG of every function in the binary.
As we limit gadgets to single paths at first and can merge them into conditional gadgets later
on in \S~\ref{gsma:ana}, we walk \emph{each} path separately.
We start our traversal from the
discovered gadget endpoints, namely ICs, IJs, and RETs. We walk every possible
path backwards until we discover a gadget starting point (EP and CS), or until
we exceed an adjustable maximum instruction length of the gadgets. The
algorithms we use are a modification of depth-first search (DFS).

First, the basic block is located containing the gadget endpoint.
Afterwards, we check if there are any calls or fixed function calls between
the endpoint and the basic block's beginning. If we encounter
a call, a CS gadget is created and the path traversal stops.
Before a gadget is added to the gadget list, we check if a gadget with
the same opcode sequence is already in that list to optionally discard or keep it
for later analysis.
If a fixed function call is
encountered, we store the information of the fixed function call and split the
current basic block. The resulting first block starts at the beginning of the
original basic block
and ends at the fixed function call. The resulting second block starts at the CS of the fixed
function call and ends with the gadget endpoint.
Thus, a CS prefixed gadget is created.
Path traversal continues and on a hit of a call, the traversal stops.
We check if the current basic block contains the EP. In that case,
we create a EP prefixed gadget.
To traverse all possible paths backwards, we keep path information and
iterate over all direct preceding basic blocks.

Then, for each block, we check if the basic block has been visited before. If that is the case, a loop gadget is only added, if
the traversed path starts at a CS and ends at a IC.
In any case, the traversal returns if the basic block has already been visited. Afterwards, the checks for
a call, fixed function call, and EP are repeated.
Finally, the instruction length  of the gadget is checked and updated.

\subsection{Gadget Analysis}
\label{gsma:ana}

Two objectives are accomplished with the gadget analysis: first, we sort out
gadgets with unsatisfiable path constraints, and second, gadgets are matched to
semantic definitions and classified accordingly. This simplifies the
utilization by a security researcher to find wanted functionality.
To make a simplified search possible, code gadgets are transformed to a symbolic
representation, executed symbolically to determine its execution contexts and
clustered into semantics due to their execution effects.

\paragraph{Lifting Code Gadgets with \emph{Zex3} to Raw Symbolic Representations.}

Code gadgets are first translated to instructions of the VEX IL. These are mapped to
Z3 expression as evaluable strings and stored offline. Thereby, most
architecture-dependent peculiarities, such as stack and flags usage, are abstracted away and implicit execution
effects are made explicit. The goal of this part of the framework, which we
named \emph{Zex3}, is to gather raw symbolic expression which are closely
related to the structure of VEX IL instructions. Thus, registers and memory
accesses are still architecture dependent.

\paragraph{Unification of Raw Symbolics with \emph{Zolver3}.}

Unification of architecture-dependent registers and memory handling is done by
a developed Z3 wrapper which we named \emph{Zolver3}. The goal
is to gather symbolic expressions for each gadget to be symbolically
evaluable by \emph{one} component only, namely Z3. Therefore, symbolic
equations created by Zex3 are transformed into a generic format, such that
register usage, memory reads and writes are adjusted.
This produces a single base usable to separate symbolic representations into
semantic bins and to verify satisfiability of each code gadget. As mentioned in
\S~\ref{gadget_disco}, each gadget is a single path. Thus, symbolic execution of
overlapping gadgets can yield conditional gadgets as well.

\paragraph{Symbolic Analysis of Code Gadgets.}

It is necessary for a security researcher during exploit development to rule out
code gadgets which do not fulfill a desired functionality. We illustrate what we
name \emph{unsatisfiability} on a gadget with a fixed function call: at the time
of compilation, it is unknown if a function call during runtime will succeed.
Therefore, checks for the return value are normally inserted in the calling
function by the developer.
Depending on the return value, a different path in the control flow is taken. We might
encounter such checks in gadgets containing a fixed function call. During
exploitation we expect the fixed function call to succeed, hence, a gadget
depending on a failed fixed function call poses unsatisfiable path constraints.

With the current level of information, a researcher is only able to search
through the discovered gadgets based on their boundaries. There is no knowledge
about the gadget's effects on the state of the to-be-exploited process during runtime.
This makes an efficient search to chain gadgets cumbersome.
Therefore, the second objective is to match every register output and every
memory effect of the symbolic representation to a semantic definition.
Zolver3 provides the state of every register and every memory effect based on
the symbolic variables and input values of the registers and memory. 
We do not have to trace every instruction
of the gadget ourself, but we can treat the gadget as a black box. We send
symbolic input values in and get all modifications to the global state of the
process by the gadget based on these symbolic input values. This means that all
register and memory store output values are symbolic expressions 
of the input values. We can use these expressions to apply our semantic
definitions to the gadgets.
The process of applying the semantic definitions to the output equations is
explained as follows.

\paragraph{Semantic Definitions.}
\label{ch:def}

In the following, we present our semantic definitions. These definitions
allow the researcher, combined with the search presented in
\S~\ref{semantic_search}, to search gadgets with specific operations performed
on a specific register or memory address.
One or more definitions are assigned to each gadget, based on the operations
the gadget performs.
When a security researcher develops a code-reuse attack, the defined gadget
types are the available instruction set. Therefore, the gadget definitions must cover all necessary instructions to perform arbitrary computations. The following gadget types
are necessary to accomplish this:
\begin{itemize}
	\item MovReg: A gadget to move the content of one register to another.
	\item LoadReg: A gadget to load a specific content into a register.
	\item Arithmetic: A gadget to perform arithmetic operations between registers.
	\item LoadMem: A gadget to load the content of a specified memory area into a register.
	\item StoreMem: A gadget to store the content of a register to a specified memory area.
\end{itemize}
We add following four semantic definitions, because they represent operations which are
commonly found in gadgets. Alternatives to extend the gadget definitions
are discussed in \S~\ref{ch:discussed}.
\begin{compactitem}
    \item ArithmeticLoad: A gadget that loads the value from a specified
        memory address, performs an
        arithmetic operation on it, and stores the result to the destination
        register.
    \item ArithmeticStore: A gadget that extends a StoreMem gadget with an
        arithmetic operation
    \item NOP - No Operation: A gadget that keeps certain registers untouched.
        This is very useful during a gadget search, because untouched registers
        can be marked as static.
    \item Undefined: If none of the previous semantic definitions match the
        equation of the register, the
        register gets marked as undefined.
\end{compactitem}
These gadget types are enough to create functionality containing jumps and
conditional jumps.
ROP uses the stack pointer to load the next instruction. Hence, an addition to or subtraction
from the stack pointer changes the next instruction. This way, the developer 
can jump through her ROP chain. JOP and COP often use a dispatcher gadget, like
the loop gadget, to invoke the gadgets of the chain. During the loop iteration
one register holds a pointer into the buffer containing subsequent gadgets. Instead of
the stack pointer (like in ROP), the register holding the pointer to the buffer
has to be modified for jumps. Conditional jumps, however, are more complicated as they have to be accomplished by chaining several arithmetic
operations~\cite{184481}. But a study of exploits~\cite{vupen1} reveals that jumping by manipulating the stack pointer is rarely 
used. Normally the chains just set the shellcode to executable and redirect the
control flow to the beginning of the shellcode. Snow et
al.~\cite{Snow:2013:JCR:2497621.2498135} come to a similar conclusion regarding
the gadget definitions in their research.

\paragraph{Applying the Definitions.}
\label{sec:app_the_def}

At the end of the symbolic execution, we have an output equation for every
register and memory write. These equations consists of Z3 \emph{expression
trees}, which represent the AST of Z3 expressions. Our definitions are stored as Z3 expression
trees as well.
Thus, we can match each symbolic operation a gadget performs against our
definition and tag the gadget with one or more definitions.

We take the approach to apply our definitions to every register and get as many operations for every gadget, as
the architecture has registers. To apply the definitions to every register, we loop over all equations belonging to classifiable registers and
perform checks if the definitions match. Classifiable registers are the general
purpose registers of the architecture and the instruction pointer.
These are the registers that are usually accessible.
We try to match every memory write to definitions recursively, because memory
accesses can be nested and every new memory store adds a new layer consisting of
Z3 store operations.

\subsection{Semantic Search}
\label{semantic_search}

In the previous steps, the gadgets have been discovered by their bounds and we
have analyzed every effect the gadgets may have on the global state of
a running process.
As we want the search for the gadgets to be flexible, we perform the search on a register and memory write basis.
One can specify the type of a single register or the types, operations, and operands of many registers.
Naturally, a search with just the type of a single register yields a lot of potential gadget candidates.
In the following section, we explain methods to order the gadget candidates and to eliminate unsatisfiable
gadgets.

\paragraph{Complexity Ordering.}

We have to present the simplest gadgets first upon a search to speed up the process of the gadget chaining. To provide the gadgets in a decreasing
complexity order, we apply four criteria.
The first criteria is that the gadgets with the lowest instruction count are
presented first. Gadgets with a low instruction count are usually simple, as
they typically do not perform many operations. 
The second criterion is to sort by the least amount of memory
writes.
For every unnecessary memory write, it has to be ensured that the write address is
inside a writable memory area.
Then the priority comes to contain the least amount of memory reads in the gadgets.
The reason is the same as for the memory writes. However, readable
memory areas are typically encountered more often and therefore easier to set
up. Our last ordering criterion requires as many registers as possible to
contain NOP definitions, as this limits unwanted side-effects such as
overwriting a register which is set up by a previous gadget.

\paragraph{Gadget Verification.}

Our gadgets support paths containing conditional branches. The exact analysis of the conditions can be tricky. For example, a gadget is needed to load the
value 0x12345678 from a specific memory address into a register.
The complexity ordering algorithm may return a gadget list with a LoadMem gadget ranked first 
that contains a conditional jump. The pitfall is that the jump is only taken, if the LoadMem operation loads a NULL value. This renders the gadget useless to
load the value 0x12345678.
Therefore, invalid gadgets similar to the one described above have to be sorted out. We automatically check the constraints of the gadget list with Zolver3 until a
satisfiable gadget is encountered.
A search query is specified by a researcher in the language \emph{Python}. Thereby the
start/end type and the content definition of the gadget is normally specified, as well as
the semantics and operations which the gadget has to fulfill.

\section{Evaluation}

In the following, we evaluate our prototype. More specifically, we analyze the
distribution of the different gadget types across different processor architectures,
demonstrate that we can discover enough gadgets for successful exploitation,
and compare our framework to existing tools.
We conducted all tests for our evaluation on a 64 bit Linux system running on an
Intel Xeon processor E3 with 3.3 GHz.
For CFG and disassembly creation, we use IDA~Pro, and VEX of Valgrind 3.9.0 is used for Zex3's translation process. Furthermore, we use pyvex's latest commit at the time of testing~\cite{pyvexcommit}.

\subsection{Gadget Type Distribution}
\label{_sec:gadget_dist}

For our evaluation, we analyzed the x86/AMD64 version of \emph{ieframe.dll} and
\emph{mshtml.dll} of Microsoft's Internet Explorer (IE) 8.0.7601.17514. We
selected these libraries as they are often used during exploitation of
IE~\cite{vupen1}. 
To evaluate our gadget finder on ARM, we analyzed Debian's (little-endian)
libc-2.19.so, because we expect \emph{libc} to always be loaded during
exploitation of a Linux system on ARM. All gadgets residing in libc-2.19.so are
in ARM mode. The gadget numbers presented in this section are the total number of
gadgets, including gadgets with and without conditional branches.

Table~\ref{tab_gadget_distri} summarizes the gadget start and end type
distribution. Note that the combination with the highest number of gadgets is
\mbox{CS-RET}. With CS-RET gadgets, one can execute common ROP exploits without
triggering CFI checks.
Due to the high proportion of CS-RET gadgets, the highest possibility to find
suitable gadgets for a gadget chain is searching for a ROP chain.
Our loop counts, also presented in Table~\ref{tab_gadget_distri}, are  based on our
loop definition. This means that all listed loops end with an IC and start at
the CS of the IC. The number of discovered loops can still be further increased
by implementing loops for JOP or allowing relaxed loop definitions.

\begin{table}[t!]
    \centering
    \caption[Gadget start and end type distribution.]{Number of available gadgets listed by gadget start and end type, and their corresponding discovery and analysis runtime.}
    \label{tab_gadget_distri}
    \begin{tabular}{rccccc}
                                & ieframe.dll & mshtml.dll & ieframe.dll & mshtml.dll & libc-2.19.so \\
    \midrule
    Architecture                & x86         & x86        & AMD64       & AMD64      & ARM          \\
    \midrule
    EP-IC                       & 4255        & 4245       & 4354        & 3947       & 261          \\
    EP-IJ                       & 59          & 370        & 172         & 1009       & 79           \\
    EP-RET                      & 11521       & 16723      & 10950       & 16517      & 2615         \\
    CS-IC                       & 36300       & 55225      & 38679       & 68791      & 1226         \\
    CS-IJ                       & 67          & 28         & 76          & 1365       & 240          \\
    CS-RET                      & 39382       & 71104      & 40831       & 72198      & 6029         \\
    Loops                       & 348         & 443        & 335         & 464        & 55           \\
    \midrule
    Runtime (s)                 & 12925.2     & 29058.7    & 16309.4     & 51259.8    & 4079.0       \\ 
    \bottomrule
    \end{tabular}
\end{table}

It is worth noting that all functions typically used by attackers for malicious
behavior are available, such as \emph{VirtualProtect} to set memory
to executable or writable, \emph{LoadLibrary} to load a library into the address space,
and \emph{CreateProcess} to create a process. Gadgets containing fixed function calls
are not restricted to some gadget start and end types, but are interspersed
throughout all start and end type combinations.
For the x86 and AMD64 DLLs mentioned in Table~\ref{tab_gadget_distri}, we found 982 gadgets with
hardcoded calls to functions which allocate memory,  
change memory permissions, load DLLs, or perform file I/O operations. 

\subsection{Exploiting ARM with One CFI-Resistant Gadget}

To evaluate our gadget finder on ARM, we exploit an artificial use-after-free
vulnerability. The instruction initiating our chain is an IC in ARM mode and the
first argument, stored in R0, contains a pointer to our prepared buffer. The
protection in place is similar to CCFIR. This means, IC and IJ can 
just transfer the control flow to EPs, and RETs are only allowed to return to legitimate CS. We assume that an information leak is available, which is usually the case for real-world exploits. Our gadget pool is derived from Debian's 
libc-2.19.so. All discovered gadgets are in ARM mode. The goal of the exploit is to execute \texttt{system("/bin/sh")}. 
\begin{figure}[h]
\begin{floatrow}
\ffigbox{%
    \centering
    \includegraphics[width=0.9\linewidth]{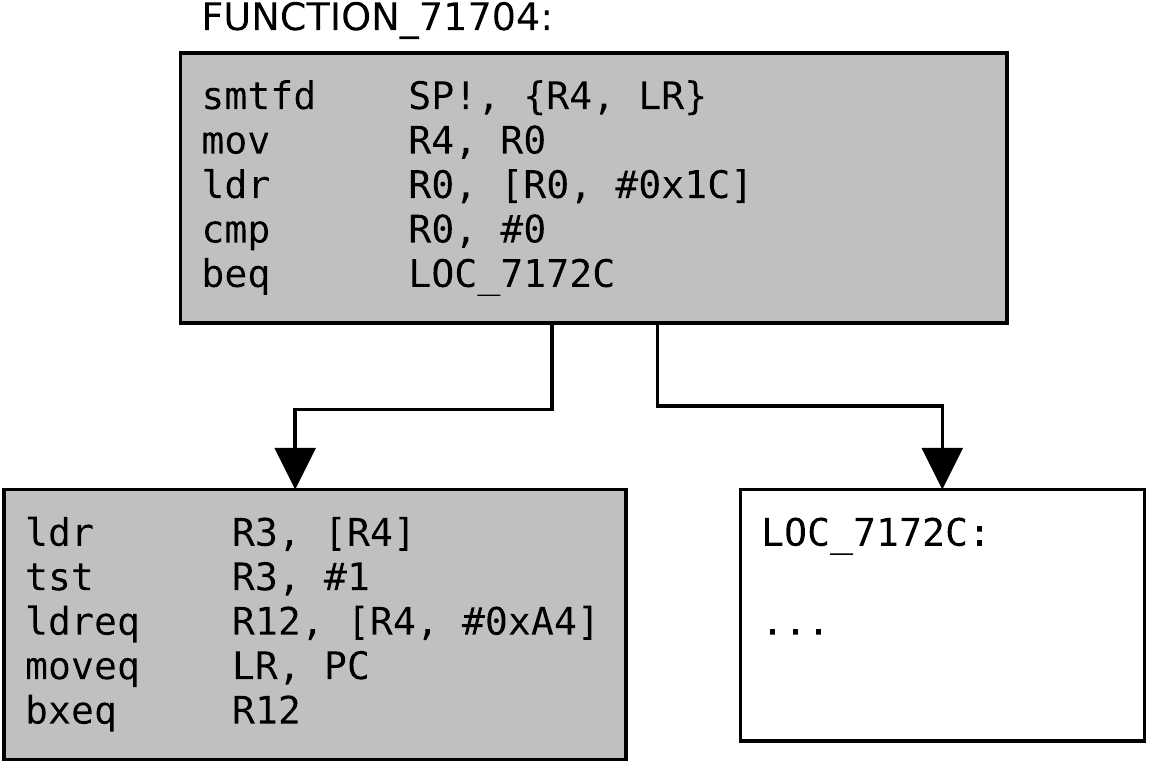}
}{%
    \caption{An ARM gadget which loads the address of \texttt{"/bin/sh"} from the supplied buffer in R0, loads the address of \texttt{system()} from the buffer to R12, and ends with an IC of R12.}
    \label{fig:gadget_arm}
}
\ffigbox{%
    \begin{flushleft}
    \ssmall
    \texttt{
    ~\\
    \# Must contain 0x00000001.\\
    Buf+0x00 =$>$ 0x00000001\\
    ...\\
    \# .rodata:00122F58 aBinSh DCB "/bin/sh",0\\
    Buf+0x1C =$>$ 0x00122F58\\
    ...\\
    \# .text:0003B190 system\\
    Buf+0xA4 =$>$ 0x0003B190\\
    ...\\
    \# Address of the first gadget\\
    \# Offset in buffer is dependent on freed object\\
    Buf+0xXX =$>$ 0x00071704\\
    }
    \end{flushleft}
}{
    \caption{Buffer exploit data. Only addresses at the offsets \texttt{0x1C} and \texttt{0xA4},
        the address for the initial control-flow transfer (\texttt{0x71704}), and \texttt{0x1} at
        offset \texttt{0x00} have to be set.}
    \label{lst:arm_buffer}
}

\end{floatrow}
\end{figure}
On ARM, the first argument to a function is not passed on the stack, but in the
register R0. Therefore, to execute \texttt{system("/bin/sh")} we have to load the
address of a string containing \texttt{"/bin/sh"} into R0. We do not have to write the
string to memory ourselves, as it is already present in libc-2.19.so. We use the
information leak to get the base address of libc-2.19.so. The address of libc-2.19.so
is also required to get the address of \texttt{system()}.
But at first, we have to find the gadgets to load the address of \texttt{system()} and the
string \texttt{"/bin/sh"}
from the buffer and call the \texttt{system()} function. These addresses are placed later on
in our buffer.
A pointer to the buffer is passed
to our gadgets in R0. Due to the protection scheme in place, the gadget has to
start at an EP. The end of the gadget is not defined, yet.
An automatically discovered gadget that exhibits the required actions is displayed in Figure~\ref{fig:gadget_arm}.
First, it loads the address of \texttt{"/bin/sh"}
from our buffer to R0 via \texttt{LDR R0, [R0,\#0x1C]}. And second, it loads the
address of \texttt{system()} to R12 and calls R12 at the end. This way, the objective to
execute \texttt{system("/bin/sh")} is achieved with a single gadget.
The buffer that we use during the exploit is shown in Figure~\ref{lst:arm_buffer}. At offset \texttt{0x00} the buffer must contain \texttt{0x1} to satisfy \texttt{TST R3,\#1}. Just if this check is valid, the address of \texttt{system()} gets loaded and called.

\subsection{Comparison to Other Gadget Discovery Tools}

To investigate how our framework performs compared to other tools, we used
ROPgadget~\cite{ROPgadget}, XROP~\cite{xrop}, and IDA
sploiter~\cite{idasploiter} to search for unique gadgets in \emph{mshtml.dll},
\emph{ieframe.dll}, and \emph{libc-2.19.so}. ROPgadget performs a semantic
search based on the disassembly of Capstone~\cite{Capstone}, while XROP and IDA
sploiter perform a standard instruction search. Thereby, IDA sploiter uses IDA~Pro. Hence, we can compare our framework to a tool which uses the same
disassembly as input.
We searched gadgets with a length of max. 30 instructions with ROPgadget and IDA
sploiter, and with a max. length of five instructions in XROP, because the
length cannot be adjusted. Then we dropped unaligned gadgets which these tools
delivered, as well as non CFI-resistant gadgets.
Overall, it is shown in Table~\ref{tab_xop_comparison} that  our tool found 1.2
times to 154.3 times more gadgets than other tools.

\begin{table}[t]
    \centering
    \tabcolsep 3pt
    \caption{Number of unique EP and CS gadgets found by other tools in comparison to
    our framework. Improvement factor states the factor of more gadgets found by our tool.}
    \label{tab_xop_comparison}
    \begin{tabular}{l l r r}
    Tool & & CFI-resistant gadgets & Improvement factor\\
    \midrule
    IDA sploiter:&libc (ARM):		&0				&ARM not supported          \\
    &ieframe.dll (x86):		&11721			&7.8\\
    &mshtml.dll (x86):			&14762			&10.0\\
    &ieframe.dll (x86\_64):		&14192			&6.7\\
    &mshtml.dll (x86\_64):		&19984			&8.2\\
    \midrule
    ROPgadget:&libc (ARM):		&8677			&1.2  \\
    &ieframe.dll (x86):		&28747			&3.2  \\
    &mshtml.dll (x86):			&30631			&4.8  \\
    &ieframe.dll (x86\_64):		&10479				&9.1 \\
    &mshtml.dll (x86\_64):		&14283				&11.5 \\
    \midrule
    XROP:&libc (ARM):		&1107			&9.4  \\
    &ieframe.dll (x86):		&660				&138.8\\
    &mshtml.dll (x86):			&957				&154.3\\
    &ieframe.dll (x86\_64):		&1531			&62.1\\
    &mshtml.dll (x86\_64):		&2479			&66.1\\
    \midrule
    Our framework: &libc (ARM):		&10450           &- \\
    &ieframe.dll (x86):		&91584           &- \\
    &mshtml.dll (x86):			&147695          &- \\
    &ieframe.dll (x86\_64):		&95062           &- \\
    &mshtml.dll (x86\_64):		&163827          &- \\
    \end{tabular}
\end{table}

\section{Related Work}

Code-reuse attacks have evolved from a simple \emph{return-into-libc}~\cite{solar} into a highly sophisticated attack vector. In times of DEP, Krahmer was the first to propose a method called \emph{borrowed code chunks} technique~\cite{krahmer}. By chaining code snippets together that end with return instructions, Krahmer showed how to perform specific operations and as a consequence bypass DEP. His work was extended by Shacham in 2007~\cite{Shacham:2007:GIF:1315245.1315313}, who showed that Turing-completeness can be achieved by reusing instruction sequences that end in return opcodes, thus leading to the name \emph
{Return-Oriented-Programming}. He called those sequences \emph{gadgets}. Large code bases typically provide enough gadgets to achieve Turing-completeness.

While the first attacks targeted the x86 architecture, the concepts have been shown to be applicable on
ARM~\cite{kornau} or SPARC~\cite{Buchanan:2008:GIG:1455770.1455776} systems as well.
ASLR~\cite{aslr} has been successful in stopping static ROP chains. However, its
ineffectiveness has also been shown in the presence of \emph{information leaks}.
Even fine-grained re-randomization can be circumvented by the means of \emph
{just-in-time ROP} as demonstrated by Snow et al.~\cite{Snow:2013:JCR:2497621.2498135}.
During the attack, they harvest gadgets based on the \emph{Galileo} algorithm
introduced by Shacham et al~\cite{Shacham:2007:GIF:1315245.1315313}. The algorithm starts at
\emph{return} instructions and iterates backwards over a code section
to retrieve gadgets that end with the return instruction.
A table lookup matches their gadgets against semantic definitions. This differs
from our approach as we lift only CFI-permitted code paths to an intermediate
representation (VEX) having a high ISA coverage, and symbolically evaluate the
gadgets to achieve a semantic binning.
Schwartz et al. developed a gadget search and compiler framework to
automatically generate ROP chains. They apply program verification techniques to
categorize gadgets into semantic definitions~\cite{schwartz2011q}. However, they do not take CFI policies into
account.

To aid in both the development of ROP attacks and CFI defenses, toolkits to
locate suitable gadgets have emerged. Frameworks such as the one introduced by
Kornau~\cite{kornau} or ROPgadget~\cite{ROPgadget} utilize an intermediate
language to abstract the underlying architecture. However, these do not locate
gadgets conforming to the constraints introduced by CFI solutions. Our
framework fills this gap and enables researchers to test their CFI policies on
multiple architectures with only one toolkit.
Closely related to our work is research which tries to measure the \emph{gadget
quality} by introducing several metrics~\cite{follner2016analyzing}. However, these metrics are bound to an
architecture, while our approach is architecture independent.

\section{Discussion}
\label{ch:discussed}

The core property of our framework is the ability to quickly test CFI policies
on multiple architectures. With the possibility to locate gadgets conforming to
the same constraints in multiple environments, we enable researches to gain a
fast overview on the security of policies. This is applicable not only to one architecture, but to all
systems supported by our toolkit. As such, it speeds up evaluation allowing
more time to be invested into the design of the the policies. The multi-platform
approach also enables to determine differences between architectures, each of which 
have an impact on the availability of certain gadget classes. 
One specific gadget class can commonly occur on one architecture, while it is nearly non-existent
on another architecture, consequently not posing a risk. Allowing researchers to
focus on the most relevant gadget classes for each architecture may lead to
defenses that fit more to the environment. 
While there are other toolkits that are able to locate gadgets on
ARM, our framework differs in that it allows to apply the same CFI policies to
different architectures. 

\paragraph{Limitations.}
At the current state, we did not include a compiler that is able to generate
complete chains from the found gadgets. While we simplify the task by
providing a query interface, the last step is still manual. The simplest
approach would be to blindly combine chains of gadgets until one of them
satisfies the constraints. However, a better solution is to
combine gadgets based on a logic that translates an intermediate language
written by a developer to a series of gadgets. However, this is no easy task as
avoding CFI detections requires longer and more complex gadgets, which are not
side-effect free. The compiler would need to account for both, the intended
effects and the compensation of any side effect of the gadget. 
Due to the modular design, we can support additional gadget types and architectures. 
For instance, it is possible to extend the discovery phase to locate unintended
instructions or whole virtual functions needed for a
COOP-attack~\cite{TUD-CS-2015-0041}. Another option is extending the
definitions by a limit of targets for an IC of a gadget. This allows assessing fine-grained CFI defenses.

\section{Conclusion}

We presented a framework that not only discovers
code-reuse gadgets across multiple architectures, but also locates gadgets that can be used
with deployed CFI defenses. While our framework can be used in an
offensive way, we deem its value for defensive research to be higher. By
quickly testing CFI constraints on multiple architectures, it is possible to
focus on the most relevant attack vectors and improve both the defensive
capabilities and the performance. In this process, we also showed that it is
possible to locate CFI-compatible gadgets not only on x86, but also on ARM.
CFI research is lacking behind on mobile platform and we hope
that by providing an effective evaluation tool, further work on this topic can be simplified.

\subsection*{Acknowledgment} 
This work was supported by ERC Starting Grant No. 640110 (BASTION).

\bibliographystyle{abbrv}
\bibliography{bibliography}
\end{document}